\documentclass[A4,12pt]{article}
\newcommand{\Ra}{\rangle}
\newcommand{\La}{\langle}

\newcommand{\Fr}{\frac{\hbar\omega}{k_BT} }

\newcommand{\Ch}{\mathop{\rm ch}}
\newcommand{\Sh}{\mathop{\rm sh}}
\newcommand{\Ct}{\mathop{\rm coth}  }

\newcommand{\E}{\mathcal E}
\usepackage{amssymb,amsmath,euscript}
\begin{document}
\begin{center}
{\large {\bf   CORRELATED COHERENT STATES - QUANTUM ANALOGUE OF THERMAL STATES}\\
 \small(ON THE PROBLEM OF INCORPORATING THERMODYNAMICS INTO QUANTUM THEORY } )\\

\large {A.D.Sukhanov$^1$,  O.N. Golubjeva$^2$, and V.G.Bar'yakhtar$^3$}\\
 \bigskip
\small {$^1$Theoretical Physics Laboratory, Joint Institute for
Nuclear Research, Dubna, Russia.  $<$ogol@oldi.ru$>$}\\
\small {$^2$People's Friendship University of Russia, Moscow,
Russia. $<$ogol@mail.ru$>$}\\
\small {$^3$ Institute of Magnetism, National Academy of Sciences of
Ukraine, Kiev, Ukraine. $<$victor.baryakhtar@gmail.com$>$}
\end{center}

\begin{abstract}
In this paper, we show that correlated coherent states (CCSs) are
the most adequate candidates for the role of quantum analogues of
the thermal states. The main result of our study reduces to the fact
that quantum thermal effects under conditions of the equilibrium of
an object with the stochastic environment at any temperatures can be
explained consistently only on the basis of the set of CCSs.
\end{abstract}

\section*{\small 1. Wave functions of correlated coherent states
of an arbitrary vacuum} In our paper [1], we proposed an original
approach to the incorporation of stochastic thermodynamics into
quantum theory. It is based on the concept of consistent inclusion
of the holistic stochastic environmental action consisting of the
thermal environmental action in addition to the quantum one.
Bogoliubov arbitrary vacua were used as a model of the generalized
thermostat. In this case, it becomes possible to bring the set of
quantum states described by the complex wave function in
correspondence with equilibrium thermal states. In the next paper
[2], it was shown that squeezed correlated states can adequately
describe only the cases of the contact with the cold vacuum, which
exists at the zero Kelvin temperature, which is not significant for
thermodynamics.

Now we consider a system (a quantum oscillator) under the conditions
of the contact with the thermostat modeled by an arbitrary vacuum.
Under these conditions, this system can be described by the complex
wave function that is dependent on the
parameter~$\alpha\ne\frac\pi2$. \footnote{The states corresponding
to~$\alpha=0$ were considered in our paper [2] and were qualified as
squeezed correlated states (SCSs), which are inadequate to the
thermal states.} In the coordinate representation, in the general
case, it has the form
\begin{equation}\label{1}
\psi_{\alpha}(q) =[2\pi (\Delta
q_0)^2\frac{1}{\cos\alpha}]^{-1/4}\exp \left\{-\frac{q^2} {4(\Delta
q_0)^2}e^{i\alpha}\right\}.
\end{equation}
Here,
\begin{equation}\label{2}
(\Delta q_0)^2 =\frac{\hbar}{2\gamma},\;\;\;\;\;\;
\end{equation}
where the coefficient~ $\gamma>0$ and $\hbar$ is the Planck
constant. The coordinate variance~ $\overline{(\Delta
q_{\alpha})^2}\equiv (\Delta q_{\alpha})^2$ in the arbitrary-vacuum
state~$|\psi_{\alpha}\Ra$  calculated using wave function (1) has
the form
\begin{equation}\label{3}
 (\Delta q_{\alpha})^2=\int_{-\infty}^{+\infty}\psi^*_\alpha q^2\psi_\alpha
dq=\frac{\hbar}{2\gamma}\cdot\frac{1}{\cos\alpha}=(\Delta
q_0)^2\cdot\frac{1}{\cos\alpha}.
\end{equation}

Similar calculations for the momentum variance~$\overline{(\Delta
p_{\alpha})^2}\equiv (\Delta p_{\alpha})^2$ in this
state~$|\psi_{\alpha}\Ra$ lead to the result
\begin{equation}\label{4}
 (\Delta p_{\alpha})^2=
\frac{\hbar\gamma}{2}\cdot\frac{1}{\cos\alpha}=(\Delta
p_0)^2\cdot\frac{1}{\cos\alpha},
\end{equation}
where $\dfrac{\hbar\gamma}{2}\equiv (\Delta p_0)^2.$

We show that wave function~(1)  is completely identical to the state
of the contact with the arbitrary vacuum~ $\psi_{\tau,\phi}$, which
was found in~[3] using the Bogoliubov $(u,v)$-transformations from
the equations

\begin{equation}\label{5-27}
\hat b\psi_{\tau,\varphi}(q,\omega)=0,
\end{equation}
where the operator~ $\hat b$ is the quasiparticle annihilation
operator for the arbitrary vacuum, or
\begin{equation}\label{6-28}
 \frac{d\psi_{\tau,\varphi}}{d
 q}+\frac{u-v}{u+v}\cdot\frac{\omega}{\hbar}\;
q\;\psi_ {\tau,\varphi}=0
\end{equation}

In the general case [4], the complex functions~$u$ and~$v$
containing in~(6) are determined in terms of the free
parameters~$(\tau,\theta,\varphi)$ as follows:
\begin{equation}\label{7-24}
u=\Ch\tau\cdot e^{i\varphi};\;\;\;\; v=\Sh\tau\cdot e^{-i\varphi}.
\end{equation}
In this case, $(\tau,\theta,\varphi)$ can be interpreted as Euler
angles, which are used to parametrize the group of rotations~$O$(3).
Without loss of generality, we assume that~$\theta=0$ in this case.

The solution of Eq.~(6) for arbitrary~ $\tau$ and~$\varphi$ has the
form of the complex Gaussoid

\begin{equation}\label{8-29}
\psi_{\tau,\varphi}(q,\omega)=C\exp\left\{-\frac{q^2}{4(\Delta
q_0)^2}\cdot\frac{u-v}{u+v}\right\},
\end{equation}
where $(\delta q_0)^2\equiv\dfrac{\hbar}{2\omega}$.

If formulas~ (7) and the normalization conditions are taken into
account, expression~ (8) becomes
\begin{equation}\label{9-30}
\psi_{\tau,\varphi}(q,\omega)=\left[2\pi(\ \Delta
q_{\tau,\varphi})^2\right]^{-1/4}\exp\left \{-\frac{q^2}{4(\Delta
q_{\tau,\varphi})^2}( 1-i\beta_ {\tau,\varphi})\right\}.
\end{equation}
Here,
\begin{equation}\label{10-31}
(\Delta q_{\tau,\varphi})^2=(\Delta q_0)^2(\Ch 2\tau-\Sh
2\tau\cdot\cos2\varphi)
\end{equation}

\begin{equation}\label{11-32}
\beta_{\tau,\varphi}=\Sh 2\tau\cdot\sin 2\varphi
\end{equation}

For the convenience of comparison of different representations ~(1)
and~(9) of the wave function, we endow them with the same forms.
Writing the exponent~$e^{i\alpha}$ in~(1) in the trigonometric form,
we obtain the result
\begin{equation}\label{12}
\psi_{\alpha}(q) =[2\pi (\Delta
q_0)^2\frac{1}{\cos\alpha}]^{-1/4}\exp \left\{-\frac{q^2} {4(\Delta
q_0)^2}\cos\alpha(1+i \tan\alpha)\right\}.
\end{equation}

To make expressions~ (1) and~(12) more similar, we assume
\begin{equation}\label{13}
\beta_{\alpha}\equiv\tan\alpha
\end{equation}
and use the value of~$(\Delta q_{\alpha})^2$ in accordance with~(3).
Finally, expression~ (1) can be written in the form

\begin{equation}\label{14}
\psi_{\alpha } =\left[2\pi(\ \Delta q_{\alpha })^2\right]^{-1/4}
\exp\left\{-\frac{q^2}{4(\Delta
q_{\alpha})^2}(1-i\beta_{\alpha})\right\}.
\end{equation}

It is natural to assume that formulas~ (1) and~(14) obtained using
different initial  preconditions and, accordingly, expressed in
terms of different parameters are nevertheless related to identical
states. Then the condition for the coincidence between~$(\Delta
q_{\alpha})^2$ (3) and~$(\Delta q_{\tau,\varphi})^2$ (10) can be
obtained if it is required that the following conditions be
satisfied at the same time:

\begin{equation}\label{15}
(\Ch 2\tau-\Sh
2\tau\cdot\cos2\varphi)\Leftrightarrow\frac{1}{\cos\alpha}
\end{equation}

\begin{equation}\label{16}
\Sh 2\tau\cdot\sin2\varphi\Leftrightarrow\tan\alpha.
\end{equation}
This turns out to be possible if in~(15) $\cos2\varphi=0$ and
in~(16) $\sin2\varphi=1$, which agrees with the well-known assertion
that the correlated coherent states (CCS) in
the~$(u,v)$-transformations are fixed by the parameter~
$\varphi=\frac\pi4.$ Thus, we demonstrated that the
states~$\psi_{\alpha}(q)$~(1) and~ $\psi_{\tau,\varphi
}(q)\Big|_{\varphi=\frac\pi4}$~(14) under condition
$\tan\alpha=\sinh 2\tau$ agree well with each other.

\section*{\small 2. Correlated coherent states as thermal ones }
To study the possibility of endowing the states~$|\psi_{\alpha}\Ra$
with the meaning of thermal states, it is necessary to bring the
parameter~$ \alpha $ in correspondence with the temperature that has
no pre-image in quantum mechanics. To do this, we consider the
expression for the Planck energy
\begin{equation}\label{17}
\E_{Pl}=\frac{\hbar\omega}{2}\coth\Fr.
\end{equation}
We note that this formula is strictly equilibrium and corresponds to
the Kelvin temperature~$T$. Therefore, based on formula~(17), we can
bring the parameter~$\alpha\ne\frac\pi2,0 $ in correspondence with
the temperature explicitly.  In what follows, we use the fact that
the average values of the kinetic~$\overline K $ and potential~
$\overline U $ energies are equal for a quantum oscillator in the
thermal-equilibrium state, so that from formula~(17), we obtain
\begin{equation}\label{18}
\overline K = \overline U
=\frac{\E_{Pl}}{2}=\frac{\hbar\omega}{4}\coth\Fr.
\end{equation}
Taking into account that $\overline K= \cfrac{1}{2m}(\overline
{\Delta p)^2}$ and $\overline U= \cfrac{m\omega^2}{2}(\overline
{\Delta q)^2}$ and assuming that $m=1$, from formula~ (18), we
obtain the coordinate and momentum variances, letting the
subscript~$T$ denote their relation with the Planck distribution:
\begin{equation}\label{19}
(\Delta q_{\scriptscriptstyle T})^2=
\frac{\hbar}{2\omega}\Ct\Fr=(\Delta q_0)^2\Ct\Fr.
\end{equation}

\begin{equation}\label{20}
(\Delta p_{\scriptscriptstyle T})^2= \frac{\hbar\omega}{2}\Ct\Fr.
\end{equation}
As follows from formulas~ (19) and~(20),  as $T$ increases, the
coordinate and momentum variances increase synchronously for thermal
states (unlike the SCSs), so that their product~ $(\mathcal
U\mathcal P)_{pq}$ also increases in this case.

The qualitative distinction between the thermal-like SCSs ($\alpha
=0$) and the thermal states (CCS) can be demonstrated obviously by
giving them the geometric interpretation in the phase plane. For
convenience, we put $\hbar=1$  and choose the dimensionless
variables~$\hat{\mathcal P}=\dfrac {\scriptstyle 1}
i\dfrac{d}{d\mathcal Q};\hat{\mathcal Q}$. Then, for cold-vacuum
states (SCSs of the~$|\psi_0\Ra$~type), the quantities~$\Delta
\mathcal P_0 $ and~$\Delta \mathcal Q_0 $ (in view of their equality
in accordance with formulas~(3) and~(4)) have the meaning of the
sides of a single square with area of $1/4$. At the same time, the
thermal CCSs (of the $|\psi_{\scriptscriptstyle T}\Ra$~type) in this
plane correspond to different squares whose sides and areas increase
consistently with temperature.

We now compare the coordinate~$(\Delta q_\alpha)^2$ (3) and
momentum~ $(\Delta p_\alpha)^2$ (4) variances calculated using wave
function~(1) with their corresponding values~ $(\Delta
q_{\scriptscriptstyle T})^2$ (19) and~$(\Delta p_{\scriptscriptstyle
T})^2$ (20) obtained using the Planck energy. They turn out to be
completely identical if we set

\begin{equation}\label{21}
\gamma=\omega
\end{equation}
and
\begin{equation}\label{22}
\frac{1}{\cos\alpha}=\Ct\Fr.
\end{equation}
Thus, in accordance with relation~ (22), the parameter~$\alpha$ in
formula~(1) fixes the states corresponding to the equilibrium at the
temperature~$T$.

Expression~(22) allows representing the exponent~$e^{i\alpha}$ in
formula~(1) in the trigonometric form, explicitly indicating its
relation with the temperature~ $T$ in this case. Calculating
$\sin\alpha=\sqrt{1-\tanh^2\Fr}=\dfrac{1}{\cosh\Fr}$ in advance, we
obtain

\begin{equation}\label{23}
e^{i\alpha}=\cos\alpha+i\sin\alpha=
\tanh(\Fr)+i\frac{1}{\cosh(\Fr)}=\tanh(\Fr)\left[1+i\frac{1}{\sinh\Fr}
\right]
\end{equation}
Returning to wave function~ (1), we can demonstrate the explicit
temperature dependence of its amplitude and phase by labeling it
with the subscript~$T$
\begin{equation}\label{24}
\psi_{\scriptscriptstyle T}(q) =[2\pi (\Delta q_0)^2
\Ct\Fr]^{-1/4}\exp \left\{-\frac{q^2} {4(\Delta
q_0)^2}\tanh(\Fr)[1+i\frac{1}{\sinh\Fr} ] \right\}.
\end{equation}

\section*{\small 3. Saturation of the Schr\"odinger UR in the correlated coherent states}
It is interesting to analyze certain CCS peculiarities. As is well
known, the most general \emph{Schr\"odinger uncertainties relation}
(SUR) for the coordinate and momentum has the form
\begin{equation}\label{25}
\Delta p\cdot\Delta q \geqslant\big|\La\psi|\delta {\hat
p}\cdot\delta {\hat q}|\psi\Ra\big|.
\end{equation}
Here, the left-hand side of the inequality contains the product of
the momentum and coordinate uncertainties  in the state~ $|\psi\Ra$
calculated using the definition
\begin{equation}\label{26}
(\Delta p)^2 \equiv\La\psi|(\delta {\hat
p})^2|\psi\Ra;\;\;\;\;(\Delta q)^2 \equiv\La\psi|(\delta {\hat
q})^2|\psi\Ra.
\end{equation}
The expression in the right-hand side
\begin{equation}\label{27}  \Big|\La\psi|\delta {\hat p}\cdot\delta {\hat
q}|\psi\Ra\Big|= \Big|\La\delta p|\delta q\Ra\Big|
\end{equation}
has the meaning of the correlator of object momentum and coordinate
fluctuations in the same state, which is expressed in terms of the
fluctuation operators~ $\delta\hat p$ and~$\delta\hat q$. We recall
that the left- and right-hand sides of relation~(22) must be
calculated independently.

In the arbitrary-vacuum state, the equality for the means

\begin{equation}\label{28}
(\Delta p_{\alpha})^2=\gamma^2(\Delta q_{\alpha})^2
\end{equation}
is valid [5] so that the \emph{left-hand} side of SUR~(25) becomes
equal to

\begin{equation}\label{29}
\Delta p_{\alpha}\cdot\Delta q_{\alpha} =\gamma\Delta q_{\alpha}^2.
\end{equation}
Calculating the correlator in the \emph{right-hand} side of SUR~(25)
with respect to the state~$ |\psi_{\alpha}\Ra$, we obtain
\begin{equation}\label{30}
\Big|\La\psi_{\alpha}|\hat p\cdot\hat
q|\psi_{\alpha}\Ra\Big|=\sqrt{\frac{\hbar^2}{4}\tan^2\alpha+\frac{\hbar^2}{4}}
=\frac\hbar2\cdot\frac{1}{\cos\alpha}.
\end{equation}
We note that the correlator here is expressed in terms of the
$\alpha$ - phase of the wave function. Taking into account that in
accordance with~ (3), $\dfrac{1}{\cos\alpha}= \dfrac 2\hbar
\cdot\gamma\Delta q_{\alpha}^2, $ we reduce correlator~ (30) to the
final form
\begin{equation}\label{31}
\Big|\La\psi_{\alpha}|\hat p\cdot\hat
q|\psi_{\alpha}\Ra\Big|=\gamma\Delta q_{\alpha}^2,
\end{equation}
which completely coincides with product~(29) of uncertainties, i.e.,
with the left-hand side of the SUR (25).

Thus, we see that the correlated state~$|\psi_{\alpha} \Ra$ is
outlined by the fact that the SUR in it indeed acquires the form of
the~\emph{equality}, i.e., becomes saturated
\begin{equation}\label{32}
\Delta p_{\alpha}\cdot\Delta q_{\alpha}=\Big|\La\psi_{\alpha}|\hat
p\cdot\hat q|\psi_{\alpha}\Ra\Big|.
\end{equation}

The interpretation of the radicand in~ (30) can be related to that
of a similar expression for the cold-vacuum state~ $\psi_0$.
For~$\alpha=0$, saturated SUR~(30) transforms into the saturated
Heisenberg  UR
\begin{equation}\label{33-24-13}
\Delta p_0\cdot\Delta q_0=\big|\La\psi_0|\frac 12[\hat p,\hat
q]|\psi_0\Ra\big|=\frac\hbar2.
\end{equation}
Here, as is known, $\dfrac\hbar2 $ is the measure of a purely
quantum environmental action occurring in the cold vacuum. Thus,
for~$ \alpha=0$, the correlator~ $\big|\La \psi_{\alpha}|\hat
p\cdot\hat q|\psi_{\alpha}\Ra\big|$ has a value that is as minimum
as possible. As was expected, the state~$|\psi_0\Ra$ indeed has the
meaning of the state of equilibrium with the cold vacuum because it
corresponds to the minimum value of the vacuum
energy~$\dfrac{\hbar\omega}{2}$.

It is adopted to regard the Heisenberg relation as a fundamental
equality reflecting the presence of unavoidable purely quantum
effects in the Nature. We assume that in comparison with~ (33), the
origin of the additional term in the radicand of correlator (30) in
the form~$\dfrac{\hbar^2}{4}\tan^2\alpha$ is related to precisely
the inclusion of the thermal influence of the arbitrary vacuum,
which is manifested in the complex character of the wave function.
The fact that the macroparameter of the effective quantum-thermostat
action

\begin{equation}\label{34-25-14}
\mathbb J_{\alpha}\equiv\sqrt{\frac{\hbar^2}{4}\tan^2\alpha+ \mathbb
J_0^2} =\frac\hbar2\frac{1}{\cos\alpha} ,
\end{equation}
(here,  $\mathbb J_0^2=\dfrac\hbar2$ relates to the purely quantum
influence) which was introduced previously in the framework of
stochastic thermodynamics [6], coincides with the right-hand side of
formula~(30) is evidence in favor of this assertion. This fact gives
grounds to assume from now on that for the arbitrary vacuum,
saturated SUR~(32) also corresponds to the more general equilibrium
state~$|\psi_{\alpha}\Ra$ occurring at the simultaneous presence of
the quantum and thermal actions. Thus, among the functions providing
the saturation of SUR~(32), there exist the functions~
$\psi_{\alpha}(q)\Big |_{\alpha\ne 0,\frac\pi2}$, which allow taking
into account extra thermal effects in addition to the quantum ones
in a certain temperature range. Thus, the CCSs can be regarded as
quantum analogues of thermal states in a sufficiently substantiated
way.

\bigskip

\small References

[1] A.D. Sukhanov and O.N. Golubjeva. Toward a quantum
generalization of equilibrium statistical thermodynamics:
$(\hbar,k)$- dynamics. Theoretical and Mathematical Physics, 160(2):
1177 (2009)

[2] A.D. Sukhanov,  O.N. Golubjeva, V.G. Bar'yakhtar.
arXiv:1211.3017v1 [quant-ph] 13 Nov 2012

[3] A.D. Sukhanov and O.N. Golubjeva. Arbitrary vacuum as a model of
stochastic  influence of environment: on the problem of
incorporating   thermodynamics into quantum theory.   Physics of
Particles and Nuclei Letters,   Vol. 9, No. 3, p. 303.
 Pleiades Publishing, Ltd., 2012.

[4]  V.V. Dodonov and  V.I.Man'ko. Trudy Fiz. Inst. Lebedev, 1987.
183, 71

[5] J.von Neumann. Mathematical Foundations of Quantum Mechanics.
Princeton University Press,1996

[6] Thermodynamics. Ed. Tadashi Mizutani, 2011.  InTech. P.P. 73-98.
\end{document}